\documentclass[aps,prd,showpacs,superscriptaddress,nofootinbib,notitlepage]{revtex4-1}
\usepackage{epsfig}
\usepackage{amsmath}
\newcommand{\ben}{\begin{eqnarray}}
\newcommand{\een}{\end{eqnarray}}
\newcommand{\nnu}{\nonumber\\}
\newcommand{\bef}{\begin{figure}[htb]\centering}
\newcommand{\eef}{\end{figure}}

\begin{document}
\title{Sivers asymmetry of Drell-Yan production in small-$x$ regime}

\author{Zhong-Bo Kang}
\affiliation{Theoretical Division, 
                   Los Alamos National Laboratory, 
                   Los Alamos, NM 87545, USA}
                   
\author{Bo-Wen Xiao}
\affiliation{Institute of Particle Physics, 
                   Central China Normal University, 
                   Wuhan 430079, China} 

\begin{abstract}
We study the Sivers single spin asymmetry of the Drell-Yan lepton pair production in the small-$x$ regime. We find that in the corresponding kinematic region the spin asymmetry calculated in the small-$x$ approach is consistent with either the usual transverse-momentum-dependent factorization formalism or the collinear factorization formalism, respectively. We estimate the Sivers asymmetry for both polarized p+p and p+A collisions and argue that the Drell-Yan production is an interesting and unique probe for both the transverse spin physics and the small-$x$ saturation effect.  
\end{abstract}

\date{\today}
\maketitle                   

\section{Introduction}
Physics at forward rapidities at Relativistic Heavy-Ion Collider (RHIC) has attracted great attention from both experimental and theoretical sides in recent years. It provides great opportunities to study novel hadronic physics phenomena, among which the single transverse spin asymmetry (SSA) in polarized p+p collisions \cite{Adams:2003fx}, and the small-$x$ gluon saturation in p+A (d+A) collisions \cite{Arsene:2004ux} are particularly interesting.
On one hand, it has been realized that the SSAs observed in high energy collisions are directly connected to the transverse motion of partons inside the polarized hadron. Thus it could help map out the three-dimensional image of the hadron in the transverse momentum space \cite{Boer:2011fh}. On the other hand, the single hadron suppression and the dihadron correlation in the forward d+Au collisions have provided a way to study the small-$x$ gluon saturation or multiple parton scattering effect 
\cite{Marquet:2007vb,Albacete:2010pg,Albacete:2010bs,Stasto:2011ru,Qiu:2004da}. An effective theory of the color-glass-condensate (CGC) has been applied to describe these phenomena. Seemingly completely different research topics, the SSAs and the small-$x$ gluon saturation can be closely related to each other and be studied in the same process \cite{Kang:2011ni}. 

The SSAs have been observed in many experiments at different collision energies. It has been found that the SSAs become the largest in the forward direction of the polarized hadron \cite{Adams:2003fx}, where the incoming partons from the polarized hadron have relatively large momentum fraction $x$, thus mostly valence quarks. At the same time, the partons in the target have very small $x$ in high energy collisions, and thus are dominated by the gluons and could be described by the small-$x$ gluon saturation physics. 
In other words, the polarized p$^\uparrow$+p and p$^\uparrow$+A in the forward rapidity region at high energy collisions will be a unique opportunity to probe both the transverse spin and small-$x$ saturation physics.

The Drell-Yan production in high energy polarized p$^\uparrow$+p and p$^\uparrow$+A collisions will be very important in such a direction. The Drell-Yan production at small transverse momentum can be described by a well-established transverse momentum dependent (TMD) factorization formalism \cite{Collins:1984kg,Ji:2004xq,Collins:2004nx,Arnold:2008kf} and thus the experimental data can be clearly analyzed and be cleanly interpreted. 
From the small-$x$ perspective, because of the lack of both final state interactions and fragmentation effects, Drell-Yan lepton pair production in p+A collisions is a very clean probe for the unintegrated gluon distribution, particularly the so-called dipole gluon distribution at small-$x$. From the spin physics perspective, Drell-Yan lepton pair production at small transverse momentum of the pair provides a way to study the so-called transverse momentum dependent distributions (TMDs) in the polarized hadron, encoding important information about the hadron structure beyond what has been learned from the usual collinear parton distribution functions (PDFs). The particularly interesting one is the quark Sivers function \cite{Siv90}, which represents a distribution of unpolarized quarks in a transversely polarized nucleon, through a correlation between the quark's transverse momentum and the nucleon polarization vector. Unlike the collinear PDFs which are universal, the TMDs or the Sivers functions are not exactly universal \cite{Bomhof:2004aw}, instead, they have the time-reversal modified universality. It was shown from the parity and time-reversal invariance of QCD that the quark Sivers function in semi-inclusive deep inelastic scattering (SIDIS) and that in Drell-Yan process are exactly opposite to each other \cite{Collins:2002kn, Brodsky}. The sign change of the Sivers function between the SIDIS and Drell-Yan production is a unique prediction of our current QCD TMD factorization formalism, and provides a critical test of the TMD factorization approach and our understanding of the SSAs \cite{Kang:2011hk}.

The quark Sivers functions have been measured in SIDIS process at HERMES, COMPASS, and JLab experiments \cite{:2009ti,:2008dn,Qian:2011py}. Future measurements of the SSAs in Drell-Yan production have been planned \cite{Collins:2005rq,Anselmino:2009st,Kang:2009sm}, in the hope of verifying the sign change in the near future. In this paper, we study the Sivers asymmetry of the Drell-Yan production in high energy collisions, particularly in the small-$x$ regime. We first formulate both the spin-averaged and spin-dependent cross section in the small-$x$ formalism or CGC framework in Sec.~II.  We then compare our small-$x$ formalism with those obtained in the TMD factorization in Sec. III. We find that the two formalisms are consistent with each other in the so-called geometric scaling region, where the invariant mass of the lepton pair $M$ is much larger than the saturation scale $Q_s$ but saturation effects are still important. In Sec. IV, we further show that the small-$x$ formalism is also consistent with the usual collinear factorization formalism in the so-called forward limit. In Sec. V, we present our numerical predictions for the Sivers asymmetry of the Drell-Yan lepton pair production in the forward rapidity region at RHIC energy for both p+p and p+A collisions. We summarize our paper in Sec. VI.

\section{Sivers asymmetry of Drell-Yan production in small-$x$ regime}
In this section, we will study the Drell-Yan lepton pair production in both transversely polarized p$^\uparrow$+p and 
p$^\uparrow$+A collisions, 
\ben
p^\uparrow(P, s_\perp)+A(P_A) \to \left[\gamma^*\to\right] \ell^+\ell^-(q)+X,
\een 
where $P$ and $s_\perp$ is the momentum and the transverse spin vector of the polarized proton, $A$
represents either a proton or a nucleus with a momentum $P_A$, and $q$ is the momentum of the lepton pair with invariant mass $q^2\equiv M^2$. To calculate the Drell-Yan cross section in small-$x$ regime, we first study the differential cross section in the high energy scattering of a quark off a color glass condensate, $q(k)+A\to \gamma^*(q)+X$. The first step in the calculation is to obtain the light-front wave functions of the incoming quark splitting into a quark and a virtual photon, $q(k, \alpha)\to \gamma^*(q, \lambda)+q(k-q, \beta)$, with $\alpha$ and $\beta$ the helicity for the incoming and outgoing quarks, respectively. The splitting wave function in momentum space is given by 
\ben
\phi_{\alpha\beta}^\lambda(k, q) = \frac{1}{\sqrt{8(k-q)^+ k^+ q^+}} \frac{\bar{u}_\beta(k-q)\gamma_\mu \epsilon^\mu(q,\lambda) u_\alpha(k)}{(k-q)^- + q^- - k^-}
\een
where $k^\mu=[k^+, k_\perp^2/2k^+, k_\perp]$ is the momentum of the incoming quark in the light-cone component, and $q$ is the momentum of the virtual photon. For virtual photon with invariant mass $M$, we could write the momentum $q^\mu$ as
\ben
q^\mu = \left[q^+, \frac{M_\perp^2}{2q^+}, q_\perp\right]
\een
with transverse mass $M_\perp=\sqrt{q_\perp^2+M^2}$. We choose the polarization of the virtual photon as
\ben
\epsilon^\mu_T(q, \lambda) = \left[0, \frac{\epsilon_\perp^\lambda \cdot q_\perp}{q^+}, \epsilon_\perp^\lambda\right],
\qquad
\epsilon^\mu_L(q, \lambda) = \frac{1}{M}\left[q^+, \frac{q_\perp^2-M^2}{2q^+}, q_\perp\right],
\een
where $\epsilon^\mu_T$ and $\epsilon^\mu_L$ are the transverse and longitudinal polarization vectors, and $\epsilon_\perp^{\lambda=1,2} = (\mp1, -i)/\sqrt{2}$. Our choice for the transverse polarization is consistent with those in Ref.~\cite{Marquet:2007vb}. With our choice for the longitudinal polarization vector, one could easily verify that the polarization vector satisfies the requirements:
\ben
q\cdot \epsilon(q, \lambda) = 0,
\qquad
\sum_\lambda  \epsilon^\mu(q, \lambda) \epsilon^{*\nu}(q, \lambda) = -g^{\mu\nu} + \frac{q^\mu q^\nu}{M^2}.
\een
The cross section can usually be written as a compact form in the transverse coordinate space, in which the splitting wave function $\psi_{\alpha\beta}^\lambda(k, q^+, r)$ is defined as the following:
\ben
\psi_{\alpha\beta}^\lambda(k, q^+, r) = \int d^2 q_\perp e^{iq_\perp\cdot r} \phi_{\alpha\beta}^\lambda(k, q).
\een
With the polarization vectors defined above, we could easily calculate $\psi_{\alpha\beta}^\lambda(k, q^+, r)$. The result is given by
\ben
\psi^{T\, \lambda}_{\alpha\beta}(k, q^+, r) &=&
2\pi \sqrt{\frac{2}{q^+}} e^{iz k_{\perp}\cdot r}
i \epsilon_M K_1( \epsilon_M |r|)
\begin{cases}
\frac{r \cdot \epsilon^{1}_\perp}{|r|}
\left[\delta_{\alpha-}\delta_{\beta-}+(1-z)\delta_{\alpha+}\delta_{\beta+}\right], 
& \lambda=1, 
\\
\\
\frac{r \cdot \epsilon^{2}_\perp}{|r|}
\left[\delta_{\alpha+}\delta_{\beta+}+(1-z)\delta_{\alpha-}\delta_{\beta-}\right], 
& \lambda=2.
\end{cases}  
\label{tran} 
\\
\psi^L_{\alpha\beta}(k, q^+, r) &=&
2 \pi \sqrt{\frac{2}{q^+}} e^{iz k_{\perp}\cdot r}
(1-z)M K_0(\epsilon_M |r| ) \delta_{\alpha\beta},
\label{long}
\een
where $z=q^+/k^+$ and $\epsilon_M^2=(1-z)M^2$, and $K_{0,1}(\epsilon_M |r|)$ are the modified Bessel functions of the second kind. It is important to realize that we have kept the transverse momentum $k_\perp$-dependence for the incoming quark, which is crucial as the spin effect comes in as a quark TMD distribution inside the polarized proton. If we set $k_\perp=0$, these splitting wave functions reduce to those derived in Ref.~\cite{Dominguez:2011br} where a collinear incoming quark distribution was considered.

\bef
\psfig{file=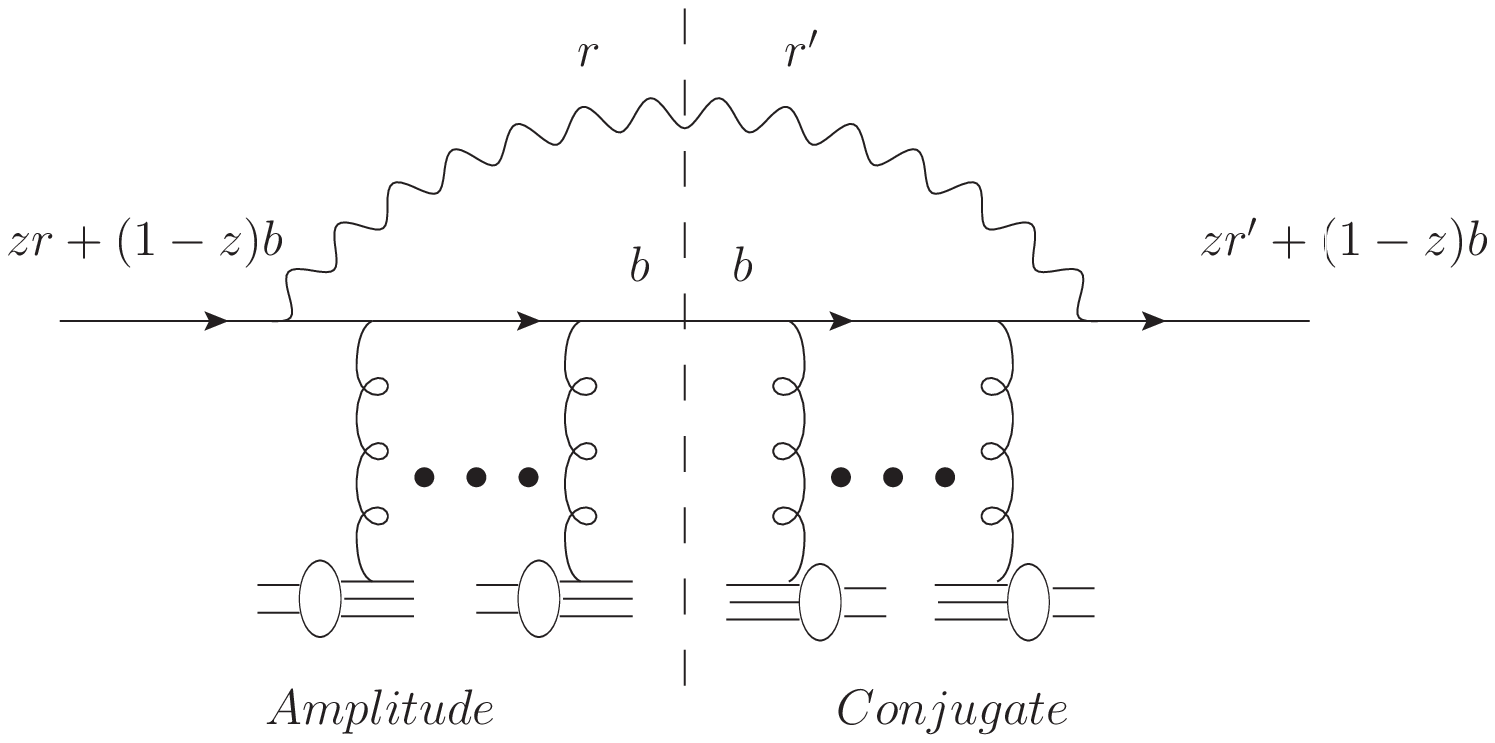, width=3.1in}
\caption{An example diagram to illustrate the interaction between the quark and the target. Interactions before and after the splitting have to be taken into account for both the amplitude and the conjugate amplitude.}
\label{scatter}
\eef
Following Ref.~\cite{Marquet:2007vb,Bjorken:1970ah,Kovner:2001vi}, when high-energy partons scatter off the CGC, the interaction is eikonal in that the projectile partons propagate through the target without changing their transverse position but picking up an eikonal phase. This phase can be expressed by the relevant Wilson lines. With the splitting wave function in the transverse coordinate space in hand, we could easily derive the differential cross section for a quark scattered off the CGC (see Fig.~\ref{scatter}), $q(k)+A\to \gamma^*(q)+X$, and the result is given by
\ben
\frac{d\sigma(qA\to \gamma^*X)}{dq^+ d^2q_\perp} &=& \alpha_{\rm em} e_q^2 
\int \frac{d^2 b}{(2\pi)^2}\frac{d^2 r}{(2\pi)^2} \frac{d^2 r'}{(2\pi)^2}
e^{-iq_\perp\cdot (r-r')}
\sum_{\alpha\beta \lambda} \psi_{\alpha\beta}^{*\lambda}(k, q^+, r'-b) \psi_{\alpha\beta}^\lambda(k, q^+, r-b) 
\nnu
& &
\times \left[1+S_{x_A}^{(2)}(v, v') - S_{x_A}^{(2)}(b, v') - S_{x_A}^{(2)}(v,b)\right]
\label{b-space}
\een
where $e_q$ is the quark fractional charge, $v=z r+ (1-z) b$ and $v'=z r'+ (1-z) b$, and the two-point functions of form $S^{(2)}(x, y)$ take care of the multiple scattering between the $\gamma^* q$-pair and the nucleus target, and they are characterized by the Wilson lines with the following definition
\ben
S_{x_A}^{(2)}(x, y) = \frac{1}{N_c}\left\langle {\rm Tr}\left(U(x) U^{\dagger}(y) \right)\right\rangle_{x_A},
\een
where the notation $\langle \cdots \rangle_{x_A}$ is used for the CGC average of the color charges over the nuclear wave function and $x_A$ is the smallest fraction of longitudinal momentum probed, and is determined by the kinematics. $U(x)$ is given by
\ben
U(x)  = \mathcal{P} \exp\left\{ig_s \int_{-\infty}^{+\infty} dx^+ T^c A_c^-(x^+, x_\perp)\right\},
\een
with $T^c$ the generator of color $SU(N_c)$ group in the fundamental representation, and 
${\mathcal P}$ denoting an ordering in $x^+$. 

Eq.~(\ref{b-space}) is a very compact form in the transverse coordinate space. In order to transparently incorporate spin-dependent effect and also to demonstrate the connection to the usual TMD factorization formalism for Drell-Yan production, we will transform Eq.~(\ref{b-space}) to the transverse momentum space, in which the cross section can be written in the following form
\ben
\frac{d\sigma(qA\to \gamma^*X)}{dy d^2q_\perp} = \frac{\alpha_{\rm em}}{2\pi^2} e_q^2 
\int d^2 b d^2 p_{\perp}  F(x_A, p_{\perp})
\big[H_T(q_\perp, k_{\perp}, p_{\perp}, z)+H_L(q_\perp, k_{\perp}, p_{\perp}, z)\big],
 \label{qt-space}
\een
where $y$ is the rapidity of the virtual photon, and $F(x_A, p_{\perp})$ is the unintegrated gluon distribution (or dipole gluon distribution) defined as
\ben
F(x_A, p_{\perp}) = \int \frac{d^2 r_\perp}{(2\pi)^2} e^{i p_{\perp}\cdot r_\perp} 
\frac{1}{N_c} \left\langle {\rm Tr}\left(U(0) U^\dagger(r_\perp) \right)\right\rangle_{x_A}.
\een
$H_T$ and $H_L$ represent the corresponding hard-part functions for transverse and longitudinal polarized virtual photon, respectively, and they are given by
\ben
H_T(q_\perp, k_{\perp}, p_{\perp}, z) &=&
\left[1+(1-z)^2\right] \left[\frac{q_\perp - z k_{\perp}}{(q_\perp - z k_{\perp})^2+\epsilon_M^2}-\frac{q_\perp - z k_{\perp}-z p_{\perp}}{(q_\perp - z k_{\perp}-z p_{\perp})^2+\epsilon_M^2}\right]^2,
\label{ht}
\\
H_L(q_\perp, k_{\perp}, p_{\perp}, z) &=&
 2(1-z)^2 M^2 \left[\frac{1}{(q_\perp - z k_{\perp})^2+\epsilon_M^2}-\frac{1}{(q_\perp - z k_{\perp}-z p_{\perp})^2+\epsilon_M^2}\right]^2.
 \label{hl}
\een

To obtain the differential cross section for Drell-Yan production in polarized 
p$^\uparrow$+p and p$^\uparrow$+A collisions from above formalism of $q+A\to \gamma^*+X$, one 
realizes that the unpolarized quark distribution inside the transversely polarized proton can be expanded as follows \cite{Bacchetta:2004jz}
\ben
f_{q/p^\uparrow}(x, k_{\perp}) = f_{q/p}(x, k_{\perp}^2) + \frac{\epsilon^{\alpha\beta} s_\perp^\alpha k_{\perp}^\beta}{M_p} f_{1T}^{\perp, q}(x, k_{\perp}^2),
\een
where $M_p$ is the proton mass, and $\epsilon^{\alpha\beta}$ is a two-dimensional anti-symmetric tensor with $\epsilon^{12}=1$. $f_{q/p}(x, k_{\perp}^2)$ is the spin-averaged quark distribution function of flavor $q$, while $f_{1T}^{\perp, q}(x, k_{\perp}^2) $ is the quark Sivers function. 
Thus by simply convoluting the spin-independent quark distribution function $f_{q/p}(x, k_{\perp}^2)$
with Eq.~(\ref{qt-space}), we immediately obtain the spin-averaged cross section 
$d\sigma\equiv[d\sigma(s_\perp) + d\sigma(-s_\perp)]/2$ for virtual photon production, $p^\uparrow(P, s_\perp)+A(P_A)\to \gamma^*(q) + X$,
\ben
\frac{d\sigma(p^\uparrow A\to \gamma^*X)}{dy d^2q_\perp} &=& \frac{\alpha_{\rm em}}{2\pi^2} \sum_q e_q^2 
\int_{x_p}^1 \frac{dz}{z} \, d^2 k_{\perp} x f_{q/p}(x, k_{\perp}^2) 
\int d^2 b d^2 p_{\perp}  F(x_A, p_{\perp})
\nnu
& &
\times \big[H_T(q_\perp, k_{\perp}, p_{\perp}, z)+H_L(q_\perp, k_{\perp}, p_{\perp}, z)\big],
\label{avg}
\een 
where $\sum_q$ runs over all light (anti)quark flavors, the quark momentum fraction $x = x_p / z$ with $x_p = M_\perp/\sqrt{s} e^y$, and the gluon momentum fraction in the target is given by $x_A=M_\perp/\sqrt{s} e^{-y}$. Eq.~(\ref{avg}) is consistent with the results derived before in \cite{dy-nokt}, if one sets the quark transverse momentum from the proton side $k_\perp\to 0$. In our case, since we are interested in the spin-dependent effect which involves TMDs, we generalize the formalism to include the $k_\perp$-dependence. 
Now convoluting the quark Sivers function with Eq.~(\ref{qt-space}), one obtains the spin-dependent cross section $d\Delta\sigma\equiv[d\sigma(s_\perp)-d\sigma(-s_\perp)]/2$ as
\ben
\frac{d\Delta\sigma(p^\uparrow A\to \gamma^*X)}{dy d^2q_\perp} &=& \frac{\alpha_{\rm em}}{2\pi^2} \sum_q e_q^2 
\int_{x_p}^1 \frac{dz}{z} \, d^2 k_{\perp} \frac{\epsilon^{\alpha\beta} s_\perp^\alpha k_{\perp}^\beta}{M_p} x f_{1T}^{\perp, q}(x, k_{\perp}^2) 
\int d^2b d^2 p_{\perp} F(x_A, p_{\perp})
\nnu
& &
\times \big[H_T(q_\perp, k_{\perp}, p_{\perp}, z)+H_L(q_\perp, k_{\perp}, p_{\perp}, z)\big].
\label{siv}
\een
The differential cross section for Drell-Yan dilepton production (decayed from virtual photon) can be easily deduced using the following relation,
\ben
\frac{d\sigma(p^\uparrow A\to \ell^+\ell^-X)}{dM^2 dy d^2q_\perp}  = \frac{\alpha_{\rm em}}{3\pi M^2}\frac{d\sigma(p^\uparrow A\to \gamma^*X)}{dy d^2q_\perp}.
\label{dy}
\een
From Eqs.~(\ref{avg}), (\ref{siv}) and (\ref{dy}), we obtain the conventionally defined SSA (or the Sivers asymmetry) for Drell-Yan dilepton production as follows
\ben
A_N = \left. \frac{d\Delta\sigma(p^\uparrow A\to \ell^+\ell^-X)}{dM^2 dy d^2q_\perp}  \right/ \frac{d\sigma(p^\uparrow A\to \ell^+\ell^-X)}{dM^2 dy d^2q_\perp}.
\label{an}
\een

\section{Connection to TMD factorization formalism}
In this section, we investigate the connection between the Drell-Yan differential cross section obtained above in the small-$x$ formalism and those from the TMD factorization approach. To find such a connection, we need to study the factorization property of the above differential cross section in the kinematic region where $M\gg q_\perp$. Furthermore, we assume that $M$ is also much larger than the saturation scale $Q_s$ which sets the transverse momentum $p_\perp$ of the unintegrated gluon distribution $F(x_A, p_\perp)$. In other words, we are going to study the region $M\gg q_\perp\sim k_\perp\sim p_\perp$ where the TMD factorization is supposed to hold for Drell-Yan production. 

In order to extract the leading power contribution from Eqs.~(\ref{avg}) and (\ref{siv}), we notice that the integral in both equations are dominated by the end point contribution of $z\sim 1$ where $\epsilon_M^2$ is in order of $q_\perp^2$ \cite{Marquet:2009ca,Mueller:1999wm}. Following Ref.~\cite{Marquet:2009ca}, we introduce a delta function: $\int d\hat{z} \delta(\hat{z} - 1/(1+\Lambda^2/\epsilon_M^2))=1$, where $\Lambda^2=(1- z) (q_\perp - k_\perp)^2 + z (q_\perp - k_\perp - p_\perp)^2$, and integrate out $z$ first. This delta-function can be further expanded in the limit of $q_\perp\ll M$,
\ben
\delta\left(\hat z - \frac{1}{1+\Lambda^2/\epsilon_M^2}\right)  &=& \frac{1-z}{\hat z}
\delta\left((1-\hat z) (1-z) - \frac{\hat z \Lambda^2}{M^2}\right)
\nnu
&\to& \frac{1-z}{\hat z} \left[\frac{\delta(1-z)}{(1-\hat z)_+} + \frac{\delta(1-\hat z)}{(1- z)_+} \right],
\label{delta}
\een
where a logarithmic term in the above expansion is power suppressed and has been neglected. Substituting this expansion to Eqs.~(\ref{avg}) and (\ref{siv}), we obtain the leading contribution to the differential cross section in the limit of $q_\perp\ll M$, 
\ben
\frac{d\sigma(p^\uparrow A\to \ell^+\ell^-X)}{dM^2 dy d^2q_\perp}  &=& \frac{\alpha_{\rm em}^2}{6\pi^3 M^4}\sum_q e_q^2
\int d^2k_\perp x_p f_{q/p}(x_p, k_\perp^2)
\int d\hat{z} \int d^2 b d^2 p_\perp  F(x_A, p_\perp) A(p_\perp, q_\perp-k_\perp, \hat{z}),
\label{avg-x}
\\
\frac{d\Delta\sigma(p^\uparrow A\to \ell^+\ell^-X)}{dM^2 dy d^2q_\perp}  &=& \frac{\alpha_{\rm em}^2}{6\pi^3 M^4}\sum_q e_q^2
\int d^2k_\perp \frac{\epsilon^{\alpha\beta} s_\perp^\alpha k_{\perp}^\beta}{M_p} x_p f_{1T}^{\perp, q}(x_p, k_{\perp}^2) 
\nnu
&&
\times
\int d\hat{z} \int d^2 b d^2 p_\perp  F(x_A, p_\perp) A(p_\perp, q_\perp-k_\perp, \hat{z}),
\label{siv-x}
\een
where $A(p_\perp, \ell_\perp, \hat{z})$ with $\ell_\perp=q_\perp-k_\perp$ is given by
\ben
A(p_\perp, \ell_\perp, \hat{z}) = \left[
\frac{\ell_\perp |\ell_\perp - p_\perp|}{(1-\hat z) \ell_\perp^2 + \hat{z} (\ell_\perp - p_\perp)^2}
-\frac{\ell_\perp - p_\perp}{|\ell_\perp - p_\perp|}\right]^2.
\label{Apt}
\een
Several comments are in order at this point. First, we found that the contribution from the second delta function $\delta(1-\hat z)$ is power suppressed. To see this more clear, one can substitute $\epsilon_M^2 = \hat z \Lambda^2/(1-\hat z)$ for both hard-function $H_T$ and $H_L$ in Eqs.~(\ref{avg}) and (\ref{siv}), there is an overall factor $(1-\hat{z})^2$. Second, we have also found that the contribution from hard-function $H_L$ (corresponding to longitudinal polarized virtual photon) is power suppressed compared to $H_T$ (corresponding to transverse polarized virtual photon). So the final leading power contributions in Eqs.~(\ref{avg-x}) and (\ref{siv-x}) are purely connected to the transverse polarized virtual photon. This is consistent with the expectation of the TMD factorization formalism, in which the virtual photon is generated from $q\bar{q}$ annihilation process and thus is transversely polarized.

On the other hand, in the TMD factorization \cite{Collins:1984kg,Ji:2004xq,Collins:2004nx,Arnold:2008kf}, we have the spin-averaged cross section as
\ben
\frac{d\sigma(p^\uparrow A\to \ell^+\ell^-X)}{dM^2 dy d^2q_\perp}  &=& \frac{4\pi\alpha_{\rm em}^2}{3N_c M^4}\sum_q e_q^2
\int d^2k_\perp d^2\ell_\perp d^2\lambda_\perp 
\delta^2\left(k_\perp+\ell_\perp+\lambda_\perp - q_\perp \right)
\nnu
&&\times
x_p f_{q/p}(x_p, k_\perp^2) x_A f_{\bar{q}/A}(x_A, \ell_\perp^2) H(M^2, x_p, x_A) S(\lambda_\perp),
\label{tmd-avg}
\een
and the spin-dependent cross section as
\ben
\frac{d\Delta\sigma(p^\uparrow A\to \ell^+\ell^-X)}{dM^2 dy d^2q_\perp}  &=& \frac{4\pi\alpha_{\rm em}^2}{3N_c M^4}\sum_q e_q^2
\int d^2k_\perp d^2\ell_\perp d^2\lambda_\perp \delta^2\left(k_\perp+\ell_\perp+\lambda_\perp - q_\perp \right)
\nnu
&&\times
 \frac{\epsilon^{\alpha\beta} s_\perp^\alpha k_{\perp}^\beta}{M_p} x_p f_{1T}^{\perp, q}(x_p, k_{\perp}^2) 
x_A f_{\bar{q}/A}(x_A, \ell_\perp^2) 
H(M^2, x_p, x_A) S(\lambda_\perp).
\label{tmd-siv}
\een
Here $f_{1T}^{\perp, q}(x_p, k_{\perp}^2)$, $f_{q/p}(x_p, k_\perp^2)$ and $f_{\bar{q}/A}(x_A, \ell_\perp^2)$ are the TMD quark Sivers function, quark and anti-quark distribution, respectively. $H(M^2, x_p, x_A)$ and $S(\lambda_\perp)$ are the hard and soft factors with the lowest order expressions:
\ben
H(M^2, x_p, x_A)=1,
\qquad
S(\lambda_\perp)=\delta^2(\lambda_\perp).
\label{soft-hard}
\een
\bef
\psfig{file=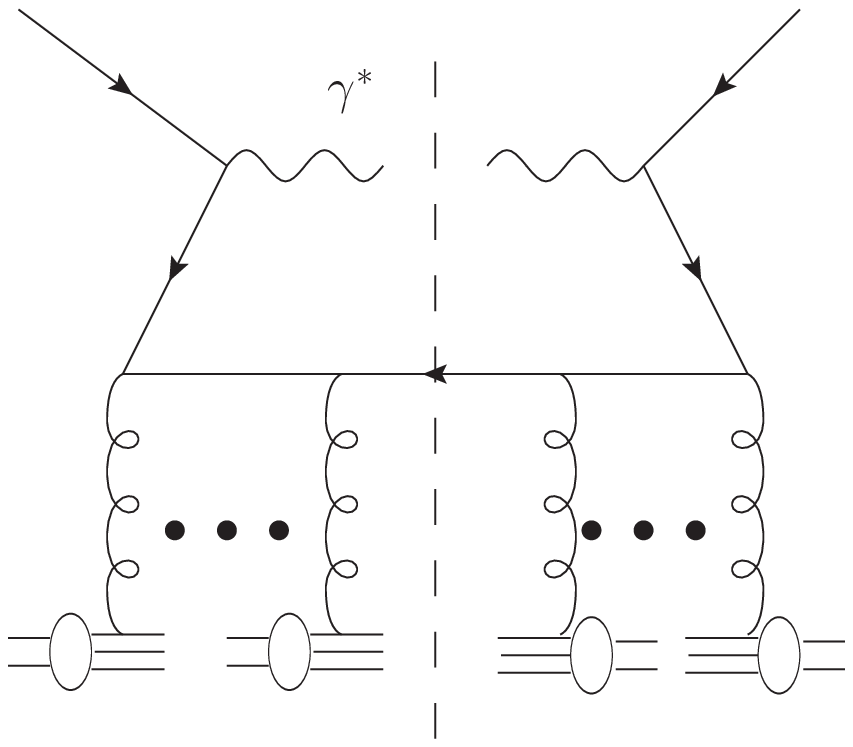, width=1.8in}
\caption{The Drell-Yan production in TMD factorization with the anti-quark distribution generated from the dipole gluon distribution.}
\label{tmd-fig}
\eef
To find the connection between the above TMD formalism and the small-$x$ formalism, we can calculate the (anti)-quark distribution function $f_{\bar{q}/A}(x_A, \ell_\perp^2)$ in terms of the dipole gluon distribution $F(x_A, p_\perp)$ at small-$x$ limit. This has been done in \cite{Marquet:2009ca}, and in the small-$x$ limit, the result can be written as the following
\ben
f_{\bar{q}/A}(x_A, \ell_\perp^2) = \frac{N_c}{8\pi^4}\int \frac{d\hat{z}}{x_A} \int d^2 b d^2 p_\perp  F(x_A, p_\perp) A(p_\perp, \ell_\perp, \hat{z})
\label{quark-kt}
\een
where $A(p_\perp, \ell_\perp, \hat{z})$ is given by Eq.~(\ref{Apt}). At the same time, since now the TMD (anti)-quark distribution starts with nontrivial leading order expansion, we could set both the hard and soft factors as their corresponding leading order expressions in Eq.~(\ref{soft-hard}). Now feed the expression for the TMD quark distribution in Eq.~(\ref{quark-kt}) back to the TMD cross sections in Eqs.~(\ref{tmd-avg}) and (\ref{tmd-siv}), taking the advantage of the delta function $\delta^2\left(k_\perp+\ell_\perp+\lambda_\perp - q_\perp \right)$ to integrate out $d^2\ell_\perp$, thus the factor $A(p_\perp, \ell_\perp, \hat{z})$ becomes
$A(p_\perp, q_\perp-k_\perp, \hat{z})$ which also appeares in the small-$x$ formalism in Eqs.~(\ref{avg-x}) and (\ref{siv-x}). We immediately find that both the spin-averaged and spin-dependent cross sections agree completely with those in Eqs.~(\ref{avg-x}) and (\ref{siv-x}) from the small-$x$ formalism. For an intuitive understanding, see Fig.~\ref{tmd-fig} where we show the Drell-Yan production in TMD factorization but with the anti-quark distribution generated from the dipole gluon distribution.
Therefore, we have demonstrated that the small-$x$ calculation for the Drell-Yan differential cross section is consistent with the TMD factorization at this particular order. 

\section{Connection to collinear factorization formalism}
In this section, we discuss the connection between the differential cross section obtained 
in the small-$x$ formalism and those in the usual collinear factorization. For this purpose, we are interested in the kinematic region where $M^2,\, q_\perp^2 \gg Q_s^2\sim p_\perp^2, \, k_\perp^2$. This is the region where the parton density in the nucleus target is still dilute and $M^2, \, q^2_\perp\gg \Lambda_{\rm QCD}$ thus the usual collinear factorization is supposed to hold.

\subsection{Spin-averaged cross section}
We will start with the spin-averaged cross section. Since we are interested in the region where $M^2,\, q_\perp^2 \gg k_\perp^2$, thus we could drop the $k_\perp$-dependence (compared to $q_\perp$) in the hard-part functions $H_{T}$ and $H_L$ in Eq.~(\ref{avg}). Once $k_\perp$ is dropped, the only $k_\perp$ dependence is then coming from the $f_{q/p}(x, k_\perp^2)$, and we could integrate over $d^2k_\perp$ to obtain the usual collinear parton distribution funciton, 
\ben
\int d^2k_\perp f_{q/p}(x, k_\perp^2) = f_{q/p}(x).
\een
On the other hand, for the nucleus target side, in the dilute parton region, we have \cite{Baier:2004tj}
\ben
\int d^2 b \, d^2p_\perp p_{\perp}^2 F(x_A, p_\perp) = \frac{2\pi^2\alpha_s}{N_c} x_A f_{g/A}(x_A),
\een
where $f_{g/A}(x_A)$ is the usual collinear gluon distribution inside the nucleus. In order to obtain the needed $p_\perp^2$ such that we could connect to the collinear gluon distribution from Eq.~(\ref{avg}), we thus need to make the expansion with respect to $p_\perp$ in the hard-part functions $H_T$ and $H_L$ in the region where $M^2,\, q_\perp^2 \gg Q_s^2\sim p_\perp^2$,
\ben
\int d^2b\int d^2 p_\perp F(x_A, p_\perp) p_\perp^\rho p_\perp^\sigma \frac{1}{2} \frac{\partial}{\partial p_\perp^\rho \partial p_\perp^\sigma}\left[H_T(q_\perp, k_\perp=0, p_\perp, z)+H_L(q_\perp, k_\perp=0, p_\perp, z)\right]_{p_\perp\to 0}.
\een
After the $p_\perp$ expansion, we set $p_\perp\to 0$, thus the hard-part functions are now independent of $p_\perp$. We could then use
\ben
\int d^2b\, d^2 p_\perp F(x_A, p_\perp) p_\perp^\rho p_\perp^\sigma = \int d^2b\,d^2 p_\perp F(x_A, p_\perp) p_\perp^2 \frac{1}{2}g_\perp^{\rho\sigma} = \frac{2\pi^2\alpha_s}{N_c} x_A f_{g/A}(x_A) \frac{1}{2}g_\perp^{\rho\sigma}.
\label{pt-int}
\een
Eventually we could write the spin-averaged cross section as
\ben
\frac{d\sigma(p^\uparrow A\to \ell^+\ell^- X)}{dM^2dy d^2q_\perp} &=& \frac{\alpha^2_{\rm em}\alpha_s}{3\pi N_c M^2} \sum_q e_q^2 
\int_{x_p}^1 \frac{dz}{z} \, x f_{q/p}(x) x_A f_{g/A}(x_A) H(q_\perp, z),
\label{small-coll-avg}
\een
where the hard-part function $H(q_\perp, z)$ is given by
\ben
H(q_\perp, z) =\frac{1}{4}g_\perp^{\rho\sigma} \frac{\partial}{\partial p_\perp^\rho \partial p_\perp^\sigma}\left[H_T(q_\perp, k_\perp=0, p_\perp, z)+H_L(q_\perp, k_\perp=0, p_\perp, z)\right]_{p_\perp\to 0}
\een
With the $H_T$ and $H_L$ given in Eqs.~(\ref{ht}) and (\ref{hl}), we could immediately derive $H(q_\perp, z)$ and it is given by
\ben
H(q_\perp, z) = \frac{z^2}{\left(q_\perp^2+\epsilon_M^2\right)^2} \left\{
\left[1+(1-z)^2\right] - \frac{2z^2q_\perp^2 \epsilon_M^2}{\left(q_\perp^2+\epsilon_M^2\right)^2}
\right\}.
\een
On the other hand, in the usual collinear factorization, the spin-averaged Drell-Yan cross section at small-$x$ regime is dominantly by the $qg\to \gamma^* q$ channel and is given by~\cite{Ji:2006vf},
\ben
\frac{d\sigma(p^\uparrow A\to \ell^+\ell^- X)}{dM^2 dy d^2q_\perp} &=& \sigma_0\frac{\alpha_s}{4\pi^2}
\sum_q e_q^2 \int \frac{dx}{x} \frac{dx_A}{x_A} f_{q/p}(x) f_{g/A}(x_A) \hat\sigma_{qg}(\hat s, \hat t, \hat u) \delta\left(\hat s+\hat t+ \hat u-M^2\right),
\label{coll-avg}
\een
where $\sigma_0=4\pi\alpha_{\rm em}^2/3N_c s M^2$, $s=(P+P_A)^2$, and $\hat\sigma_{qg}(\hat s, \hat t, \hat u) $ has the following form
\ben
 \hat\sigma_{qg}(\hat s, \hat t, \hat u) =2\,T_R\left[-\frac{\hat s}{\hat t} - \frac{\hat t}{\hat s} - \frac{2M^2 \hat u}{\hat s \hat t}\right],
\een
with color factor $T_R=1/2$. Now to connect to the small-$x$ formalism, realizing
\ben
\hat s=\frac{q_\perp^2+\epsilon_M^2}{z(1-z)},
\qquad
\hat t=-\frac{q_\perp^2+\epsilon_M^2}{z},
\qquad
\hat u=-\frac{q_\perp^2}{1-z},
\label{stu}
\een
we obtain
\ben
 \hat\sigma_{qg}(\hat s, \hat t, \hat u) = \frac{1}{1-z}  \left\{
\left[1+(1-z)^2\right] - \frac{2z^2q_\perp^2 \epsilon_M^2}{\left(q_\perp^2+\epsilon_M^2\right)^2}
\right\}.
\een
Then plugging above equation back to Eq.~(\ref{coll-avg}), integrating over $dx_A$ with the delta function $\delta(\hat s+\hat t+\hat u-M^2)$ and using $dx/x=dz/z$, we immediately end up with an expression exactly the same as the one in Eq.~(\ref{small-coll-avg}). Thus we have found that the small-$x$ formalism is exactly the same as the usual collinear factorization formalism for the spin-averaged cross section in the relevant kinematic region.

\subsection{Spin-dependent cross section}
Now let us study whether there is some connection between the spin-dependent cross section $d\Delta\sigma/dM^2 dyd^2q_\perp$ in the samll-$x$ formalism and that derived in the collinear factorization approach. Again we are interested in the region where  $M^2,\, q_\perp^2 \gg Q_s^2\sim p_\perp^2, \, k_\perp^2$. In the small-$x$ formalism, it is given by Eq.~(\ref{siv}). For the polarized proton side, we have the $k_{\perp}^{\beta} f_{1T}^{\perp,q}(x, k_\perp^2)$, thus we need another linear $k_\perp$ term from the hard-part functions $H_T$ and $H_L$ to have the $d^2k_\perp$ integral nonvanishing. On the other hand, for the unpolarized nucleus side, to obtain the usual collinear gluon distribution function, we need to perform the expansion with respect to $p_\perp$ just like what we have done for the spin-averaged case. In total, we should perform the following expansion,
\ben
\frac{d\Delta\sigma(p^\uparrow A\to \ell^+\ell^-X)}{dM^2 dy d^2q_\perp} &=& \frac{\alpha^2_{\rm em}}{6\pi^3 M^2} \sum_q e_q^2 
\int_{x_p}^1 \frac{dz}{z} \, d^2 k_{\perp} \frac{\epsilon^{\alpha\beta} s_\perp^\alpha k_{\perp}^\beta}{M_p} x f_{1T}^{\perp, q}(x, k_{\perp}^2) 
\int d^2b d^2 p_{\perp} F(x_A, p_{\perp})
\nnu
& &
\times 
k_\perp^\gamma p_\perp^\rho p_\perp^\sigma
\frac{1}{2}\frac{\partial}{\partial k_\perp^\gamma \partial p_\perp^\rho \partial p_\perp^\sigma}
\big[H_T(q_\perp, k_{\perp}, p_{\perp}, z)+H_L(q_\perp, k_{\perp}, p_{\perp}, z)\big]_{k_\perp\to 0, p_\perp\to 0}.
\een
Then we can perform the $d^2k_\perp$ integral as follows:
\ben
\frac{1}{M_p}\int d^2 k_\perp k_\perp^\beta k_\perp^\gamma f_{1T}^{\perp,q}(x, k_\perp^2)
=\frac{1}{M_p}\int d^2 k_\perp k_\perp^2 \frac{1}{2} g_\perp^{\beta\gamma} f_{1T}^{\perp,q}(x, k_\perp^2)
= \frac{1}{2} g_\perp^{\beta\gamma} T_{q, F}(x, x),
\een
where we have used the relation between the quark Sivers function $f_{1T}^{\perp,q}(x, k_\perp^2)$ and the twist-3 quark-gluon correlation function $T_{q, F}(x, x)$ \cite{Kang:2011hk}
\ben
\frac{1}{M_p}\int d^2 k_\perp k_\perp^2 f_{1T}^{\perp,q}(x, k_\perp^2)=T_{q, F}(x, x).
\een
For the nucleus target side, we perform the $d^2p_\perp$ integral the same way as in Eq.~(\ref{pt-int}) for the spin-averaged cross section in last subsection. At the end of day, we have 
\ben
\frac{d\Delta\sigma(p^\uparrow A\to \ell^+\ell^-X)}{dM^2 dy d^2q_\perp} = \frac{\alpha^2_{\rm em}\alpha_s}{3\pi N_c M^2}  \epsilon^{\alpha\beta} s_\perp^\alpha \sum_q e_q^2
\int_{x_p}^1 \frac{dz}{z} x\,T_{q, F}(x, x) x_A f_{g/A}(x_A) H^\beta(q_\perp, z),
\label{coll-small-spin}
\een
where the hard-part function $H^\beta(q_\perp, z)$ is given by
\ben
H^\beta(q_\perp, z) = 
\frac{1}{8} g_\perp^{\rho\sigma}
\frac{\partial}{\partial k_\perp^\beta \partial p_\perp^\rho \partial p_\perp^\sigma}
\big[H_T(q_\perp, k_{\perp}, p_{\perp}, z)+H_L(q_\perp, k_{\perp}, p_{\perp}, z)\big]_{k_\perp\to 0, \, p_\perp\to 0}.
\een
We could easily work out the explicit expression for $H^\beta(q_\perp, z)$,
\ben
H^\beta(q_\perp, z) = q_\perp^\beta \frac{2z^3}{\left(q_\perp^2+\epsilon_M^2\right)^3}
\left\{ \left[1+(1-z)^2\right] - \frac{z^2\epsilon_M^2 (3q_\perp^2-\epsilon_M^2)}{\left(q_\perp^2+\epsilon_M^2\right)^2}\right\}.
\label{small-hbeta}
\een

On the other hand, in the collinear factorization approach, the spin-dependent $d\Delta\sigma/dM^2 dyd^2q_\perp$ in the region $M^2,\, q_\perp^2\gg \Lambda_{\rm QCD}$ is a twist-3 effect and has the following form \cite{Ji:2006vf, note}
\ben
\frac{d\Delta\sigma(p^\uparrow A\to \ell^+\ell^-X)}{dM^2 dy d^2q_\perp} &=&
\sigma_0 \epsilon^{\alpha\beta} s_\perp^\alpha q_\perp^\beta \frac{\alpha_s}{4\pi^2} \sum_q e_q^2
\int \frac{dx}{x} \frac{dx_A}{x_A} f_{g/A}(x_A) \delta\left(\hat s+\hat t+ \hat u-M^2\right)
\frac{1}{-\hat u}
\nnu
&&
\times
\Bigg\{
\left[T_{q,F}(x, x) - x\frac{d}{dx} T_{q,F}(x, x)\right] H_{qg}^s(\hat s, \hat t, \hat u)
+ T_{q,F}(x, x) N_{qg}^s(\hat s, \hat t, \hat u) 
\nnu
&&
~~~ + T_{q, F}(x-\bar{x}_g, x) H_{qg}^h(\hat s, \hat t, \hat u)
\Bigg\},
\label{coll-spin}
\een
where $\bar{x}_g = -x \hat t/(M^2-\hat t)$, both $H_{qg}^s$ and $N_{qg}^s$ are the hard-part functions associated with the so-called soft-gluon pole contribution, and $H_{qg}^h$ is the hard-part function associated with the hard-gluon pole contribution. These hard-part functions are given by \cite{Ji:2006vf, note}
\ben
H_{qg}^s(\hat s, \hat t, \hat u) &=& \frac{N_c^2}{N_c^2-1} \left[-\frac{\hat s}{\hat t} - \frac{\hat t}{\hat s} - \frac{2M^2 \hat u}{\hat s\hat t}\right],
\\
N_{qg}^s(\hat s, \hat t, \hat u) &=& \frac{N_c^2}{N_c^2-1} \frac{M^2}{\hat s\hat t^2}
\left[M^4-2M^2\hat t+\hat u^2\right],
\\
H_{qg}^h(\hat s, \hat t, \hat u) &=& \frac{\left(M^2-\hat t \right)^3+M^2\hat u^2}{\hat s \hat t^2}
\left[2T_R \frac{\hat s}{\hat s+\hat u} - \frac{N_c^2}{N_c^2-1} \right].
\een

The collinear factorization expression in Eq.~(\ref{coll-spin}) seems very different from the small-$x$ formalism in Eq.~(\ref{coll-small-spin}). However, if one considers the so-called forward limit, i.e., $|\hat t|\ll |\hat u|\sim \hat s$, we will find that they are indeed consistent with each other as we will show now. 
From Eq.~(\ref{stu}), the forward limit is equivalent to $q_\perp^2\sim \epsilon_M^2$ and $z\to 1$. In this limit, the small-$x$ formalism in Eq.~(\ref{coll-small-spin}) reduces to
\ben
\frac{d\Delta\sigma(p^\uparrow A\to \ell^+\ell^-X)}{dM^2 dy d^2q_\perp} = \frac{\alpha^2_{\rm em}\alpha_s}{3\pi N_c M^2}  \epsilon^{\alpha\beta} s_\perp^\alpha q_\perp^\beta \frac{1}{(2q_\perp^2)^3} \sum_q e_q^2
\int_{x_p}^1 dz\,  x\,T_{q, F}(x, x) x_A f_{g/A}(x_A).
\label{smallx-forward}
\een
On the other hand, for the collinear factorization expression in Eq.~(\ref{coll-spin}), we find that the hard-part function $H_{qg}^s\propto 1/(1-z)$ while both $N_{qg}^s$ and $H_{qg}^h$ are proportional to $1/(1-z)^2$. Thus $H_{qg}^s$ is sub-leading compared to $N_{qg}^s$ and $H_{qg}^h$ in the forward limit and can be neglected. At the same time, in the forward limit, $\bar{x}_g = -x\hat t/(M^2-\hat t)\to 0$ thus the hard-pole correlation function $T_{q, F}(x-\bar{x}_g, x)\to T_{q, F}(x, x)$ is equal to the soft-pole correlation function.
Once we realize this, we find that the corresponding terms $\propto N_c^2/(N_c^2-1)$ in $N_{qg}^s$ and $H_{qg}^h$ cancel each other, only the term $\propto T_R$ remains. After using the $\delta\left(\hat s+\hat t+ \hat u-M^2\right)$ to take care of the $dx_A$ integration and $dx/x=dz/z$, we end up with the following expression
\ben
\frac{d\Delta\sigma(p^\uparrow A\to \ell^+\ell^-X)}{dM^2 dy d^2q_\perp} = \frac{\alpha^2_{\rm em}\alpha_s}{3\pi N_c M^2}  \epsilon^{\alpha\beta} s_\perp^\alpha q_\perp^\beta \frac{1}{(2q_\perp^2)^3} \sum_q e_q^2
\int_{x_p}^1 dz\,  x\,T_{q, F}(x, x) x_A f_{g/A}(x_A),
\een
which is exactly the same as the forward limit expression Eq.~(\ref{smallx-forward}) in the small-$x$ formalism. Thus we have demonstrated for the spin-dependent cross section that the collinear twist-three formalism is consistent with the small-$x$ formalism in the forward limit. Since both the spin-averaged and spin-dependent cross sections are consistent with the small-$x$ formalism in the forward limit, the Sivers spin asymmetry $A_N$ as a ratio of these two cross sections is thus consistent between the small-$x$ formalism and the collinear factorization approach.

\section{Numerical results}
In this section, we present our numerical estimates for the Sivers asymmetry of Drell-Yan lepton pair production in both polarized p$^\uparrow$+p and p$^\uparrow$+A collisions at RHIC energies by using Eqs.~(\ref{avg}), (\ref{siv}), (\ref{dy}) and (\ref{an}). To be consistent with the experimental definition for the asymmetry $A_N$, we choose a frame in which the polarized proton moves in the $+z$-direction, the unpolarized proton (or the nucleus) moves in $-z$-direction, the transverse spin vector $s_\perp$ and the transverse momentum of the lepton pair $q_\perp$ along $y$- and $x$-directions, respectively. 
For the transversely polarized proton side, we use the quark Sivers function extracted from SIDIS in \cite{Anselmino:2008sga} and flip the sign for the Drell-Yan production. For spin-averaged cross section, we use GRV98LO unpolarized parton distribution functions \cite{Gluck:1998xa}. 
In the current paper, we will not incorporate the evolution effect for the Sivers function~\cite{Aybat:2011ge,Kang:2008ey} and will leave such a study for a future publication.
For the target proton or nucleus side, we need the dipole gluon distribution function $F(x_A, p_\perp)$, which characterizes the dense gluon distribution in the Drell-Yan process in the proton or nucleus at the small-$x$ regime. In the numerical estimate, we choose two parametrizations for the dipole gluon distribution: the GBW model in \cite{GolecBiernat:1998js}, and the solution of the BK evolution equation \cite{Balitsky:1995ub}.

In the GBW model, the dipole gluon distribution has a Gaussian form \cite{GolecBiernat:1998js},
\ben
F(x_A, p_\perp) = \frac{1}{\pi Q_s^2(x_A)} e^{-p_\perp^2/Q_s^2(x_A)},
\een
where the saturation scale for the proton $Q_s^2(x_A) = Q_{s0}^2 (x_A/x_0)^{-\lambda}$ with $Q_{s0}=1$ GeV, $x_0=3.04\times 10^{-4}$, and $\lambda=0.288$. For the saturation scale of the nucleus, we follow Ref.~\cite{Dusling:2009ni} and choose
\ben
Q_{sA}^2(x_A) = c\, A^{1/3} Q_{s0}^2 (x_A/x_0)^{-\lambda},
\een
with $c=0.5$ for the minimum bias p+A collisions.

\bef
\psfig{file=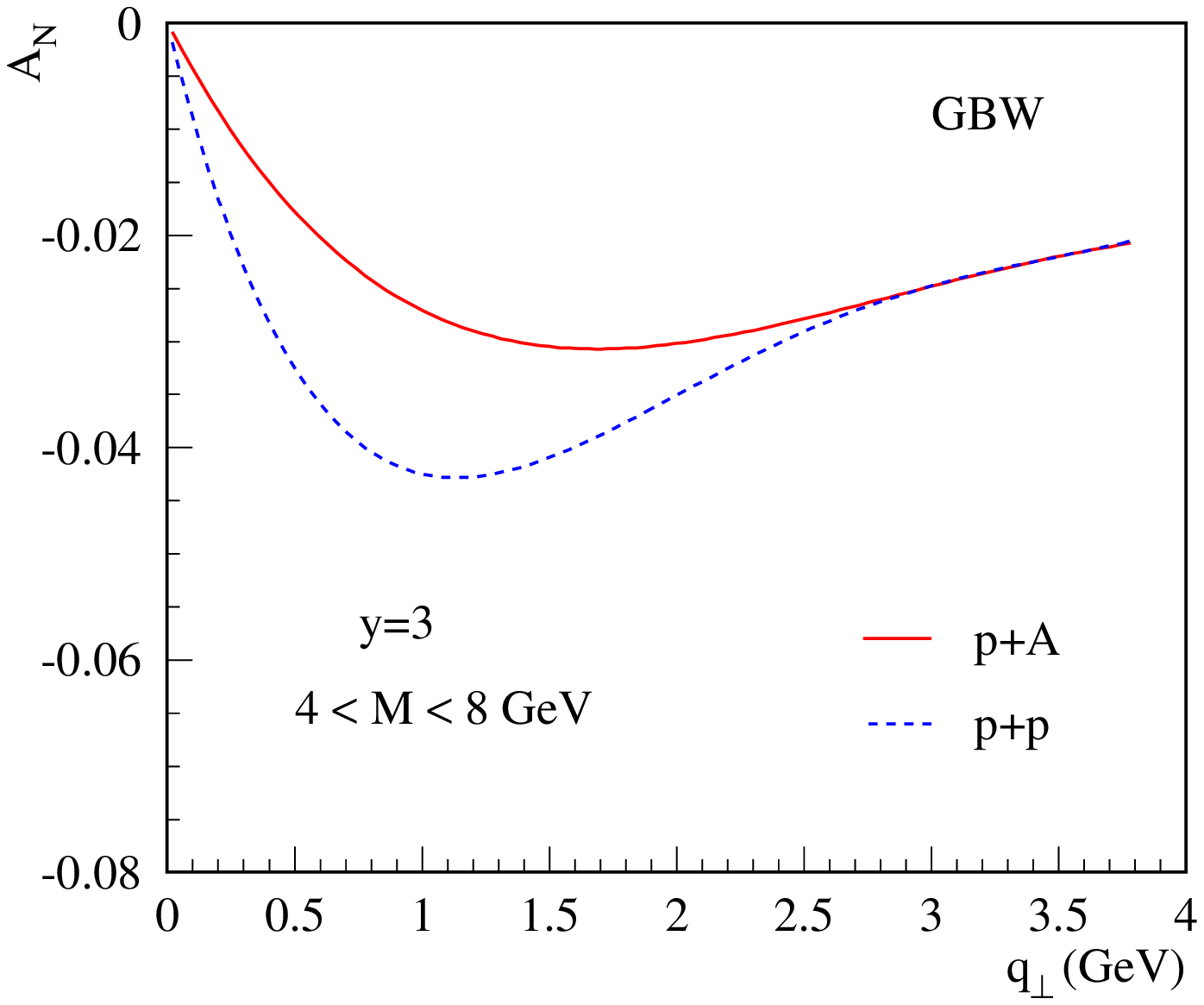, width=3in}
\hskip 0.3in
\psfig{file=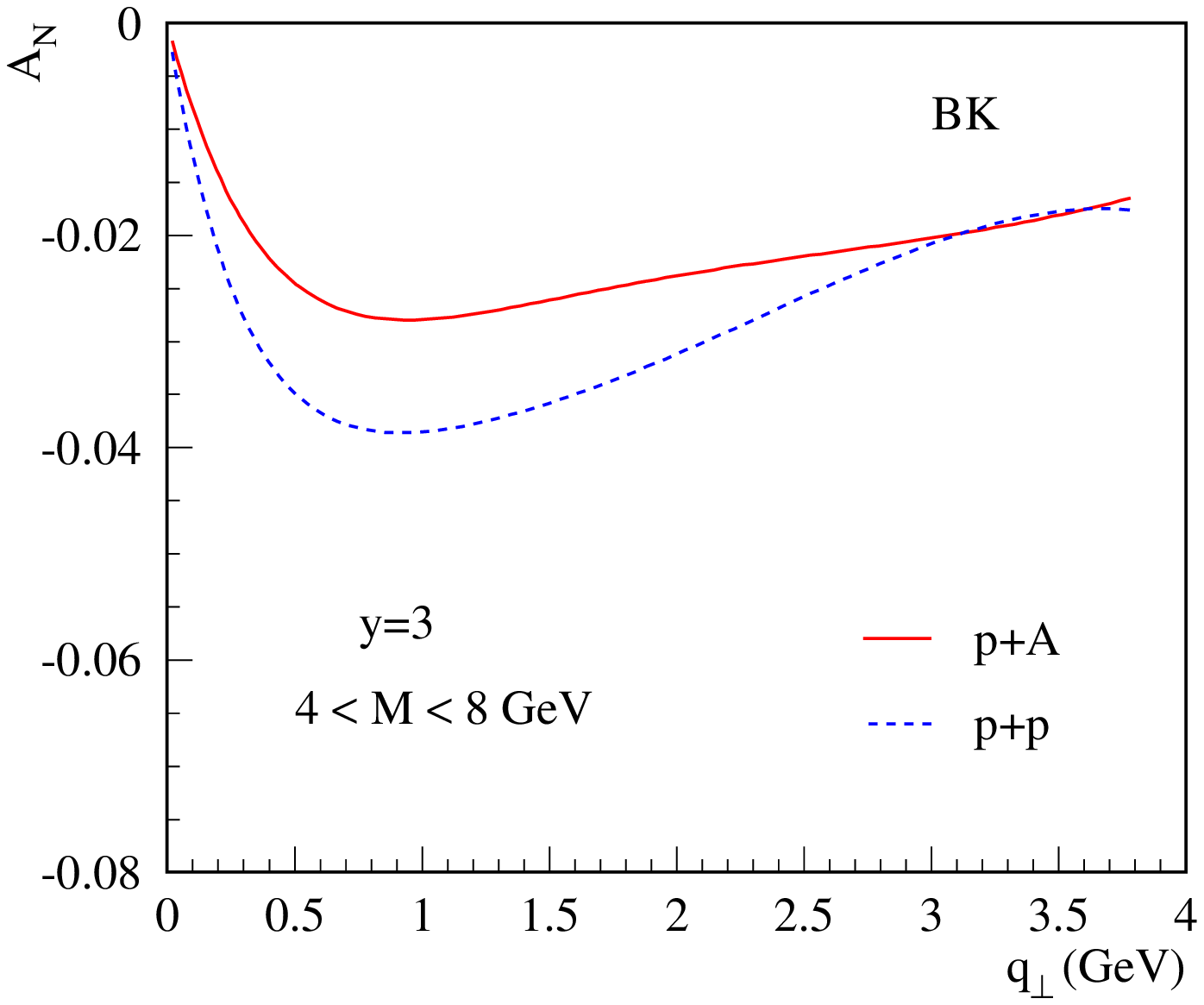, width=3in}
\caption{The single transverse spin asymmetry $A_N$ is plotted as a function of the transverse momentum $q_\perp$ at rapidity $y=3$ for RHIC energy $\sqrt{s}=510$ GeV. We have integrated over the invariant mass of the lepton pair $4<M<8$ GeV. The red solid curve is for polarized p$^\uparrow$+A collisions, while the blue dashed curve is for polarized p$^\uparrow$+p collisions. On the left plot, 
we have used GBW model for the dipole gluon distribution; while on the right plot, the dipole gluon distribution is taken as the numerical solution of the BK evolution equation.}
\label{fig-qtdep}
\eef

For the second choice, we use the numerical solution of BK evolution equation for the dipole gluon distribution in our calculation. Particularly we take those used in Ref.~\cite{Albacete:2010bs} which are able to describe successfully the single inclusive hadron production in forward d+Au collisions at RHIC
in the leading order formalism. Ref.~\cite{Albacete:2010bs} has used McLerran-Venugopalan (MV) model \cite{McLerran:1997fk} as the initial condition for the BK evolution equation, which avoids an unphysical exponential fall-off of the dipole gluon distribution at large transverse momenta. 

In Fig.~\ref{fig-qtdep}, we plot the Sivers asymmetry $A_N$ of the Drell-Yan dilepton production as a function of the pair's transverse momentum $q_\perp$ at RHIC energy $\sqrt{s}=510$ GeV. To keep in the forward rapidity region (such that the gluon momentum fraction $x_A$ in the target is small thus to justify the use of our small-$x$ formalism), we choose the rapidity $y=3$. We have integrated the pair's invariant mass $M$ from 4 to 8 GeV. On the left panel, we show the result by using the GBW model for the dipole gluon distribution. On the right panel, the asymmetry is calculated with the dipole gluon distribution taken as the numerical solution of the BK evolution equation. The blue dashed curves are for the asymmetry in polarized p$^\uparrow$+p collisions, and the red solid curves are for the asymmetry in polarized 
p$^\uparrow$+A collisions. We find that the GBW model and the numerical solution of the BK evolution equations actually give very similar results on the Sivers asymmetry for both p$^\uparrow$+p and p$^\uparrow$+A collisions. The asymmetry goes to zero when the pair's transverse momentum goes to zero $q_\perp=0$. This is because that once integrated over both $k_\perp$ and $p_\perp$ in Eq.~(\ref{siv}), the only remaining transverse vectors are $q_\perp$ and the spin $s_\perp$, both of which are needed to form the azimuthal spin asymmetry $\sim \epsilon^{\alpha\beta} s_\perp^\alpha q_\perp^\beta$. Thus if $q_\perp=0$, there are no enough vectors to generate the asymmetry. 

\bef
\psfig{file=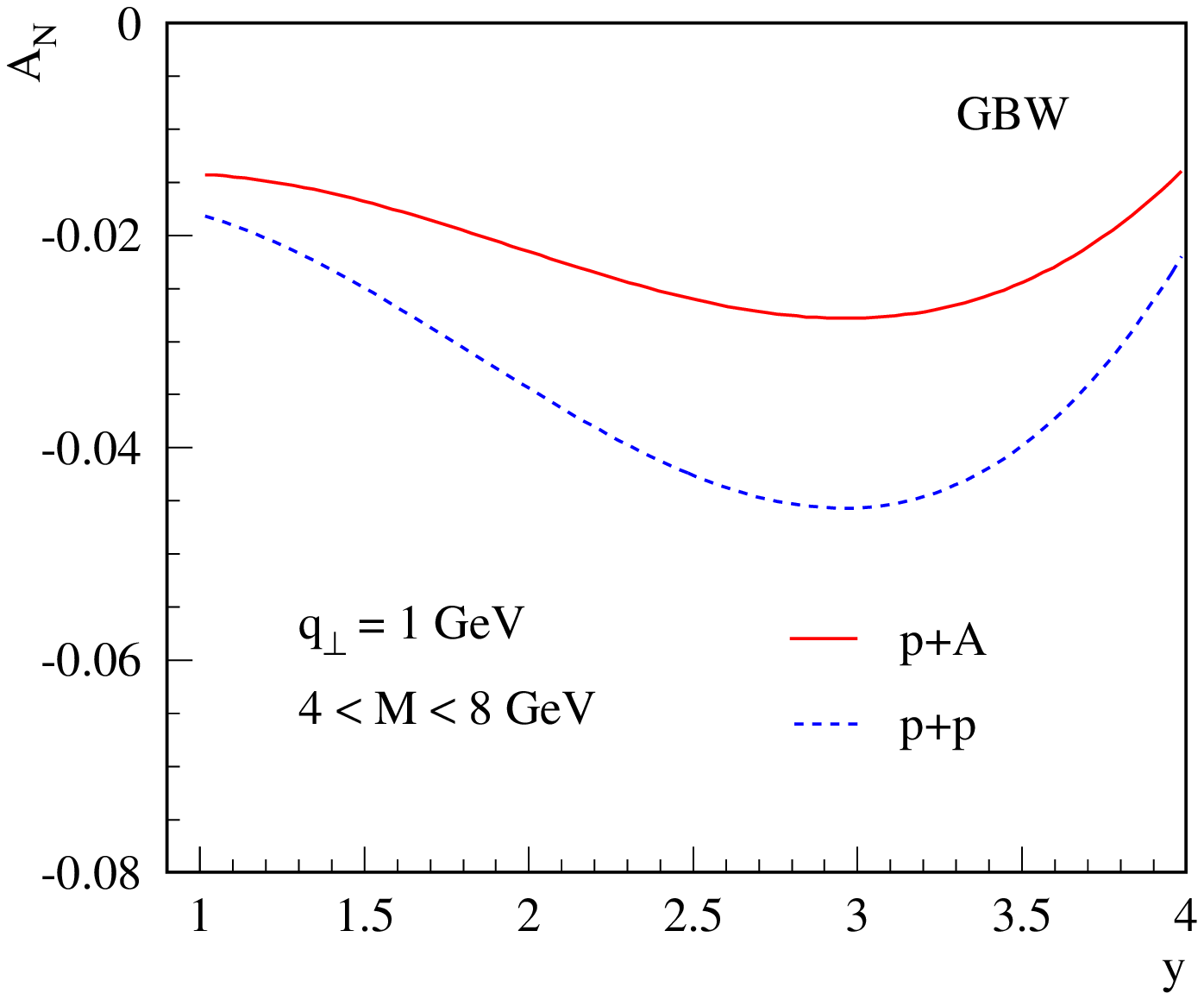, width=3in}
\hskip 0.3in
\psfig{file=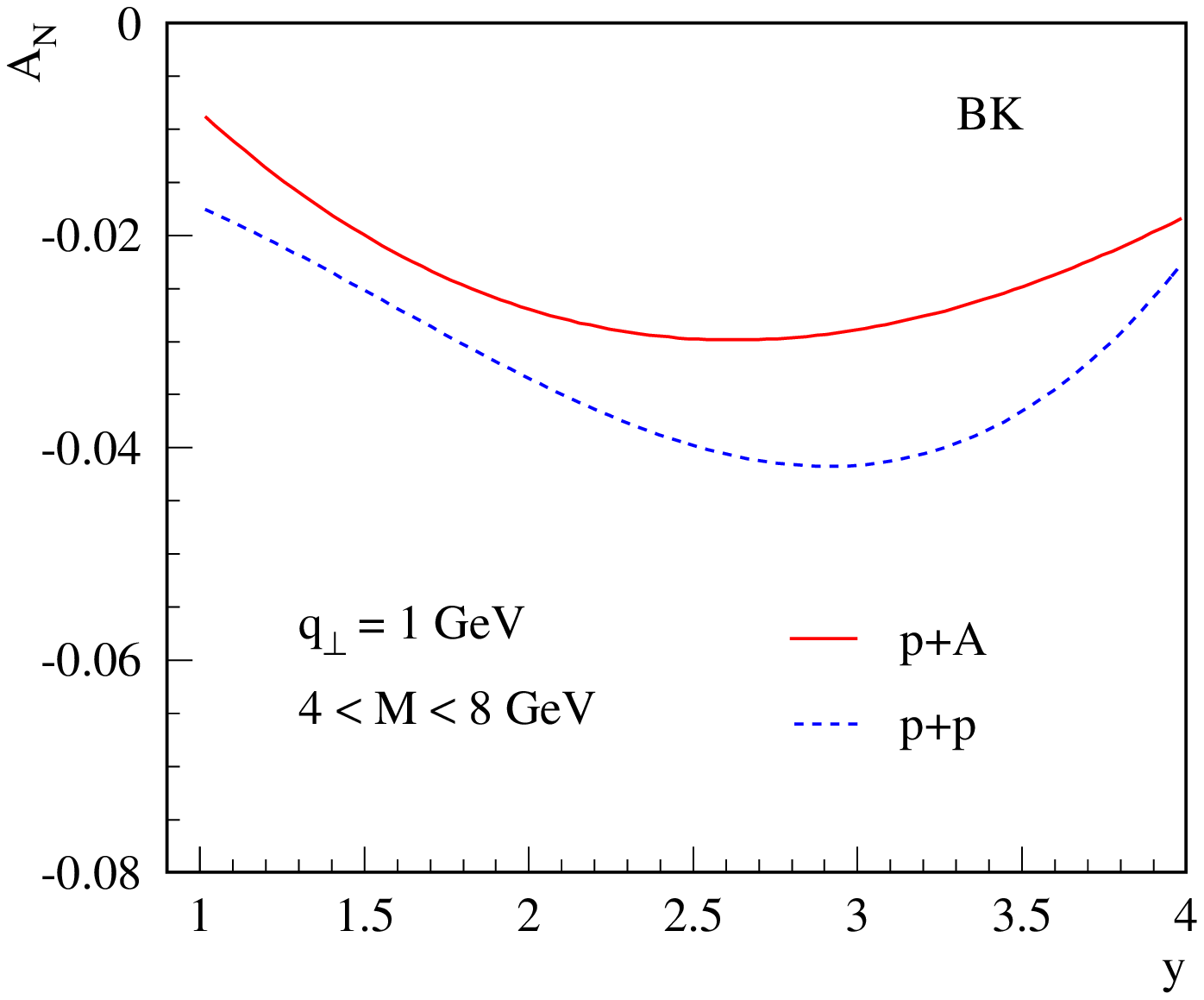, width=3in}
\caption{The single transverse spin asymmetry $A_N$ is plotted as a function of the rapidity $y$ at $q_\perp=1$ GeV for RHIC energy $\sqrt{s}=510$ GeV. We have integrated over the invariant mass of the lepton pair $4<M<8$ GeV. The red solid curve is for polarized p$^\uparrow$+A collisions, while the blue dashed curve is for polarized p$^\uparrow$+p collisions. On the left plot, 
we have used GBW model for the dipole gluon distribution; while on the right plot, the dipole gluon distribution is taken as the numerical solution of the BK evolution equation.}
\label{fig-ydep}
\eef

We also find that for relatively small $q_\perp$, the Sivers asymmetry $A_N$ becomes smaller from p$^\uparrow$+p to p$^\uparrow$+A collisions. This is expected since the Sivers asymmetry is generated by the small quark transverse momentum $k_\perp$ in the polarized proton, whereas the role that the saturation momenta play in this process is to push the produced Drell-Yan lepton pair towards larger $q_\perp$ region. Since the saturation scale $Q_s$ is larger in the nucleus target, the Sivers effect is attenuated and thus the asymmetry $A_N$ is smaller for the nucleus case as compared to the case with proton targets for fixed value of $q_\perp$. On the other hand, when $q_\perp$ becomes relatively large (larger than the saturation scale $Q_{sA}$ in the nucleus), the saturation effect (due to the difference in $Q_s$) becomes less important, thus we see the asymmetry becomes similar for p$^\uparrow$+p and p$^\uparrow$+A collisions in this region. A comparative measurement of the asymmetry for both p$^\uparrow$+p and p$^\uparrow$+A collisions thus might provide an interesting and novel way to study the saturation scale. 

In Fig.~\ref{fig-ydep}, we plot $A_N$ as a function of the dilepton rapidity $y$ at transverse momentum $q_\perp=1$ GeV at RHIC energy $\sqrt{s}=510$ GeV. The left plot is for GBW model and the right plot is for the numerical solution of the BK evolution equation case. Again $y$-dependence of the asymmetry is similar for these two cases, and the asymmetry is smaller for p$^\uparrow$+A collisions than that for p$^\uparrow$+p collisions, due to the same reason as in the $q_\perp$-dependence case.  It is easy to understand the reason why the asymmetry is peaked at $y\sim 3$, which corresponds to the parton momentum fraction in the polarized proton $x_p\sim 0.2$. This effect is simply due to the fact that the parametrization we used for the Sivers function has a maximum at $x\sim 0.2$  \cite{Anselmino:2008sga}. Nevertheless, we also find that the saturation effect plays a non-trivial role in the calculation of the asymmetry.

\section{Summary}
We studied the Sivers single spin asymmetry of the Drell-Yan lepton pair production in the forward rapidity region (small-$x$ regime) for both polarized p$^\uparrow$+p and p$^\uparrow$+A collisions. 
By including the transverse momentum dependence of the quark distribution inside the polarized proton, we calculated both the spin-averaged and spin-dependent Drell-Yan differential cross sections in the small-$x$ formalism. In the so-called geometric scaling region in which the invariant mass $M$ of the pair is much larger than the saturation scale $Q_s$ and the pair's transverse momentum $q_\perp$, we demonstrate that the Drell-Yan cross section in the small-$x$ formalism agrees completely with those derived from the transverse momentum dependent factorization approach. We further show that the small-$x$ formalism is also consistent with the collinear factorization formalism in the so-called forward limit. We evaluated the SSAs of Drell-Yan lepton pair production in both polarized p$^\uparrow$+p and p$^\uparrow$+A collisions for RHIC energy $\sqrt{s}=510$ GeV at forward rapidity region. We find that in the region when the pair's transverse momentum $q_\perp$ is small, the Sivers asymmetry is smaller in p$^\uparrow$+A collisions as compared to p$^\uparrow$+p collisions, while the asymmetry becomes similar at relatively large $q_\perp$. The future measurement and comparative study of the Sivers single spin asymmetry for the Drell-Yan lepton pair production in both polarized p$^\uparrow$+p and p$^\uparrow$+A collisions will be very interesting and important as they probe both the transverse spin and the small-$x$ saturation physics.

\section*{Acknowledgments}
We thank D. Boer, Y. Kovchegov, Y.-Q. Ma, J.-W. Qiu, R. Venugopalan, and F. Yuan for helpful discussions, and thank J. L. Albacete for providing us their unintegrated gluon distribution used in our numerical estimate. This work was supported in part by the U.S. Department of Energy under Contract No.~DE-AC52-06NA25396.

\end{document}